\documentclass[a4paper,fleqn,usenatbib]{mnras}

\usepackage[T1]{fontenc}
\usepackage{ae,aecompl}

\usepackage{graphicx}	
\usepackage{amsmath}	
\usepackage{amssymb}	
\usepackage{natbib}

\usepackage{hyperref}
\usepackage{siunitx}
\hypersetup{colorlinks,citecolor=blue,linkcolor=blue,urlcolor=blue}

\newcommand{\arcm}{\mbox{\ensuremath{\mkern-4mu^\prime}}}
\newcommand{\arcs}{\mbox{\ensuremath{\!\!^{\prime\prime}}}}

\title[Optical Counterparts of X-1 in NGC 2500]{Optical Counterparts of an Ultraluminous X-Ray Source X-1 in NGC 2500}

\author[N. Aksaker et al.]{N. Aksaker$^{1,2}$\thanks{E-mail:naksaker@cu.edu.tr}, A. Akyuz$^{2,3}$, S. Avdan$^{2}$ and H. Avdan$^{2}$
\\
$^1$Adana Organised Industrial Zones Vocational School of Technical Science, University of Cukurova, Adana, Turkey\\
$^2$Space Science and Solar Energy Research and Application Center (UZAYMER), University of Cukurova, Adana, Turkey\\
$^3$Department of Physics, University of Cukurova, Adana, Turkey
}

\date{Accepted XXX. Received YYY; in original form ZZZ}

\pubyear{2019}

\begin{document}
\label{firstpage}
\pagerange{\pageref{firstpage}--\pageref{lastpage}}
\maketitle

\begin{abstract}
We present the results of a search for optical counterparts of ultraluminous X-ray source (ULX) X-1 in the nearby galaxy NGC 2500 by using archival images taken with the Hubble Space Telescope ({\it HST}) Wide Field Camera (WFC3)/UVIS. Four optical sources have been identified as possible counterparts within the 2$\sigma$ error radius of 0$\farcs$3 in the images. However, only two of them were investigated as candidates for counterparts due to their point-like features and their identification in various filters.
These two faint candidates have absolute magnitudes of $M_{\rm V}$ $\approx$ -3.4 and -3.7. Also possible spectral classes of them were determined as B type main sequence stars. The ages and the masses of the candidates from Color Magnitude Diagram (CMD) were estimated as 45 Myr and 7 $M_{\rm \odot}$, respectively.
The spectral energy distributions (SEDs) of two candidates were modeled by a power-law spectrum with a photon index ($\alpha$) $\sim$1.5. The spectra with such slopes could be interpreted as an evidence of reprocessing of the X-rays in the outer part of the disk that generates optical emission.
\end{abstract}




\section{Introduction}
\label{sec:intro}

Ultraluminous X-ray sources (ULXs) are variable X-ray sources located away from the nucleus of galaxies. These sources emit X-rays with isotropic luminosities in excess of $10^{39}$ erg s$^{-1}$. This corresponds to approximately the Eddington limit for 10 $M_{\rm \odot}$ Black Hole (BH) (for recent review see \cite{2017ARA&A..55..303K}). There are several current models to explain the observed luminosities of this particular class of objects. One such model suggests ULXs with a stellar mass black hole $M_{\rm BH}$ $\leq$ 20 $M_{\rm \odot}$ can be powered by either a super Eddington accretion rate and/or a strong beaming or any combination of both (\citealp{2013MNRAS.432..506P}; \citealp{2009MNRAS.393L..41K}; \citealp{2014Natur.514..198M}; \citealp{2015NatPh..11..551F}). Alternatively, ULXs can also be intermediate-mass black holes (IMBHs), $10^{2}$ $M_{\rm \odot}$ $\leq$ $M_{\rm BH}$ $\leq$ $10^{4}$ $M_{\rm \odot}$, which are accreting at sub-Eddington rates (\citealp{1999ApJ...519...89C}; \citealp{2009Natur.460...73F}, \citealp{2018MNRAS.478.2576M}).

Furthermore, recent discoveries of four ULXs exhibiting coherent pulsations which are powered by neutron stars (M82 X-2 \citealp{2014Natur.514..202B}; NGC5907 ULX-1, \citealp{2017Sci...355..817I}; NGC 7793 P13, \citealp{2016ApJ...831L..14F}; \citealp{2017MNRAS.466L..48I} and NGC 300 ULX1, \citealp{2018MNRAS.476L..45C}). Hence, the debate on the true nature of ULXs is continuing.

Multiwavelength observations help us understand the nature of these binary sources. In addition to X-rays, their optical (including infrared and ultraviolet) and radio observations provide important information to constrain on masses, emission mechanisms and interactions with their environments. Determination of optical counterparts allow us to measure rotation curves of the companion stars. However, since ULXs are faint sources in optical band ($m_{\rm V}$ = 21$-$26) (\citealp{2011ApJ...737...81T}), it is difficult to measure their radial velocity curves, hence, determination of their mass functions. By making use of large and up to-date ground based and Hubble Space Telescope ({\it HST}), optical counterparts of about 20 ULXs have been identified (e.g. \citealp{2008A&A...486..151G}; \citealp{2011ApJ...733..118Y}; \citealp{2013ApJS..206...14G};
\citealp{2014Natur.514..198M}; \citealp{2016ApJ...828..105A}; \citealp{2016ApJ...831...56U}; \citealp{2018ApJ...854..176V}). Such studies give important hints to estimate the origin of optical emission, which has strong contribution from the accretion disk as well as donor star. It is also possible to obtain information	about various properties of the	donor stars (e.g. their ages, masses, and spectral types) and their environments.

In this work, we present the results of a search for optical identification of one of the ULXs (X-1) of the galaxy NGC 2500 by using the archival images of {\it HST}/WFC3/UVIS camera with various filters. NGC 2500 is a spiral galaxy (type SB(rs)d) with a low surface brightness at a distance of $\sim$10 Mpc (\citealp{1988ngc..book.....T}). Two ULXs were catalogued by \cite{2011ApJ...741...49S} in this galaxy. X-1 is located 52$^{\prime\prime}$ away from the center with a {\it Chandra} coordinates, R.A. = $08^{h}$ $01^{m}$ $48^{s}.10$, Dec. = $+$ $50^{\circ}$ 43 $\arcm$ $54\farcs60$ having an unabsorbed X-ray luminosity of $5\times$10$^{39}$ erg s$^{-1}$ in the 0.3$-$10 keV energy band. 
Besides the fact that no optical counterpart of the ULX X-1 has been identified so far, \cite{2017MNRAS.469..671L} also reported that a candidate counterpart of the source could not be detected in the near infrared. They estimated the apparent limiting magnitude as >20.16$\pm$0.06 in the H-band. Although X-2 has been classified as a ULX, \cite{2013A&A...549A..81G} identified that it is a background AGN at $z=0.5140$ by investigating its spectral features. 

The paper is organized as follows: observations and details of data analysis are described in Section \ref{sec:obs}. Discussion of the various optical properties of the ULX with a summary are given in Section \ref{sec:res}.

\section{Observations and Data Analysis}
\label{sec:obs}

\begin{figure*}
	\begin{center}
		\includegraphics[width=\textwidth]{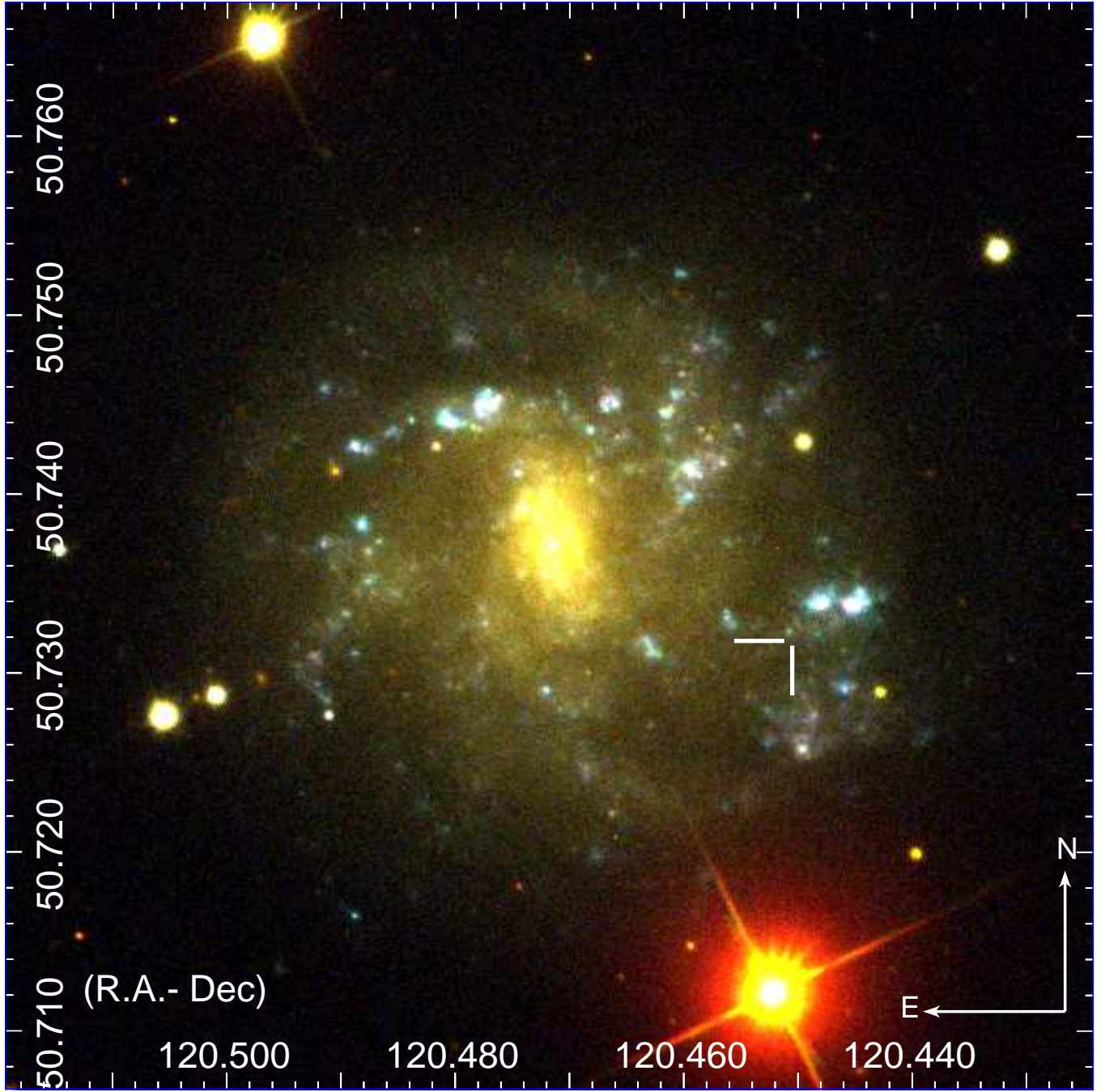}
		\caption{Three color SDSS image of NGC 2500. Red, green, and blue colors represent i, r and u bands, respectively. The white lines show the location of ULX X-1.}
		\label{Aksaker_fig1}
	\end{center}
\end{figure*}

Archival data of NGC 2500 from {\it HST}/WFC3/UVIS were utilized to search for the optical properties of ULX X-1. The observation log given in Table \ref{tab1}. The three color SDSS image of the NGC 2500 with the approximate position of ULX X-1 are shown in Figure \ref{Aksaker_fig1}.

\subsection{Astrometry and the Optical Counterparts}

\begin{figure*}
	\begin{center}
		\includegraphics[width=\textwidth]{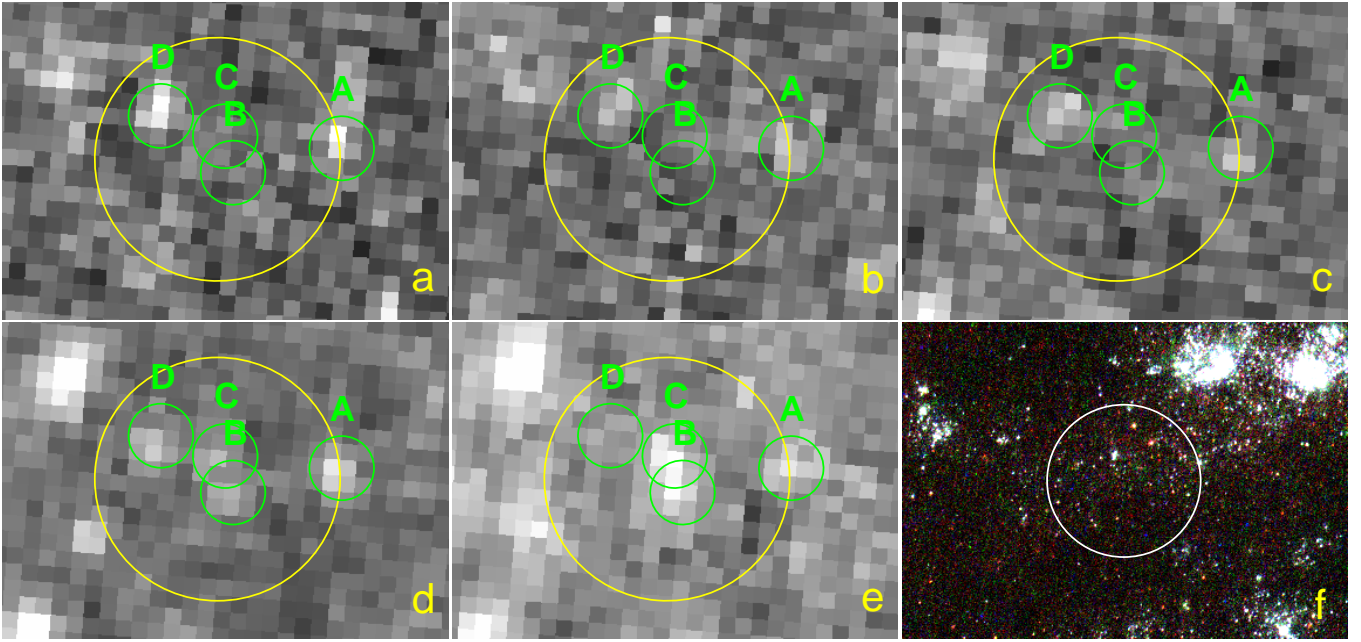}
		\caption{{\it HST}/WFC3/UVIS images of ULX X-1 (a:F275W, b:F336W, c:F438W, d:F555w and e:F814W) and the true color RGB {\it HST} image (f, red:F555W, green:F438W and blue:F336W). The yellow circle represents the corrected position of ULX X-1 with 0.3$^{\prime\prime}$ error radius. Four possible optical counterparts (green circles which are 0.12$^{\prime\prime}$ radius) are detected within the error circle. 
			The white circle (with 5$^{\prime\prime}$ radius) represents the field stars around the ULX X-1.}
		\label{Aksaker_fig2}
	\end{center}
\end{figure*}

We used the {\it HST}/WFC3 F555W image (see Table \ref{tab1}) and the {\it Chandra} image (ObsID 7112) to obtain the astrometric correction for the precise source position. The required relative astrometric corrections between images were performed by the alignment of common reference sources in both images. Source detection steps were carried out using {\it daofind} in {\scshape iraf} for {\it HST} and {\it wavdetect} task in {\scshape ciao} for {\it Chandra}. Among the detected sources, two point-like sources were adopted as the reference objects for the calculation of the relative shift between {\it HST} and {\it Chandra} images. One of the reference sources is the previously cataloged QSO (R.A. =$08^{h}$ $01^{m}$ $57^{s}.85$, Dec. = $+$ $50^{\circ}$ 43 $\arcm$ $39.\arcs80$, \citep{2015PASA...32...10F}) and the other is the center of the galaxy. These sources have X-ray counts (with statistical error radii) of 22 ($0\farcs08$) and of 7 ($0\farcs12$), respectively. The shifts between 
{\it HST} and {\it Chandra} images were derived $-0\farcs04$ for R.A and $0\farcs16$ for Dec. Then the corrected position of X-1 is determined as R.A.= $08^{h}$ $01^{m}$ $48^{s}.14$, Dec.=$+$ $50^{\circ}$ 43 $\arcm$ $54.\arcs55$ with a 2$\sigma$ positional error radius of 0$\farcs$3. Four possible optical counterparts were identified within this error circle. The corrected position of X-1 on {\it HST}/WFC3/UVIS images together with the possible optical counterparts are shown in Figure \ref{Aksaker_fig2}. These candidates are labeled as A, B, C and D according to increasing R.A. coordinates. Similar multiple optical counterparts for ULXs were also reported for various other galaxies \citep{2005ApJ...629..233K, 2008ApJ...675.1067F, 2015ApJ...812L..34W}	

\subsection{Photometry}
The Point Spread Function (PSF) photometry was performed with {\scshape dolphot} (v2.0) package with WFC3 modules. The data files (*flt.fits and *drz.fits) were retrieved from the {\it HST} data archive\footnote{https://archive.stsci.edu/hst/search.php}. Data reduction steps were followed as given in the {\scshape dolphot} manual \cite{2000PASP..112.1383D}.
The {\it wfc3mask} task was used to remove pixels flagged as bad and multiplies the pixel values by the pixel areas to convert to number of electrons on the raw images (designated *.flt or *.flc). {\it Splitgroups} and {\it calcsky} tasks were performed to split the UVIS image files into a single file for each chip of WFC3 and to create a sky image, respectively. Then {\it dolphot} task was used to detect the optical sources, their photometry, and their photometric conversion. The {\it dolphot} zero points and aperture corrections were applied to the results as default. As a result of the processes the magnitudes were calculated in the VEGA magnitude system. The galactic extinction towards NGC 2500 is $E(B-V)=0.037$ (\citealp{2011ApJ...737..103S}). The coordinates, reddening corrected VEGA magnitudes, absolute magnitudes and colors of candidates are given in Table \ref{tab2}. Candidates B and C have not been clearly detected in F275W, F336W and F438W filters, moreover, they have extended features. Hence, these candidates were not considered as possible optical counterparts. At about 70$^{\prime\prime}$ southwest of the galaxy center, there is the bright red star 2MASS J08014853+5042433. The effect of the illumination of this star is quite noticeable, especially in F555W and F814W filters. Therefore, these two filters are not used for further analyses.

\begin{figure}
	\begin{center}
		\includegraphics[width=\columnwidth]{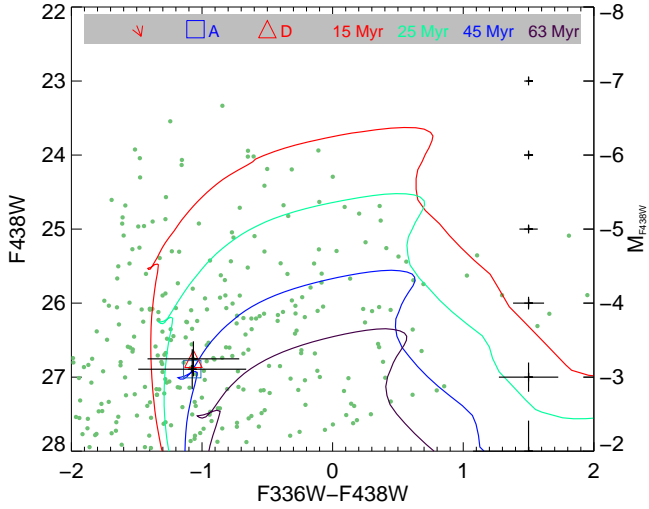}
		\caption{CMD for the possible optical counterparts (A and D) and field stars. The blue square and red triangle represent A and D, respectively. The green dots indicate field stars within 5$^{\prime\prime}$ radius around ULX X-1. The red, green, blue and brown solid lines show 15 Myr, 25 Myr, 45 Myr and 63 Myr ages, respectively. The colored isochrones have been corrected for extinction of $A_{B}=0.19$ mag and red arrow shows the reddening line. The error bars on the right indicate mean magnitude errors of field stars.}
		\label{Aksaker_fig3}
	\end{center}
\end{figure}

The Color-Magnitude Diagram (CMD) was obtained for age and mass estimation of each possible optical counterparts. For this, PARSEC isochrones were downloaded from the web\footnote{http://stev.oapd.inaf.it/cgi-bin/cmd}. These isochrones were selected with solar metallicity and {\it HST}/WFC3/UVIS wide filters photometric system. The galactic reddening and distance modulus (30.02 mag) were used to produce the CMD. The resultant CMD, F336W (U) versus F336W$-$F438W (U$-$B), for possible optical counterparts of X-1 with the field stars ($\approx$ 500) are given in Figure \ref{Aksaker_fig3}.

\subsection{Spectral Energy Distributions}
\begin{figure}
	\begin{center}
		\includegraphics[width=\columnwidth]{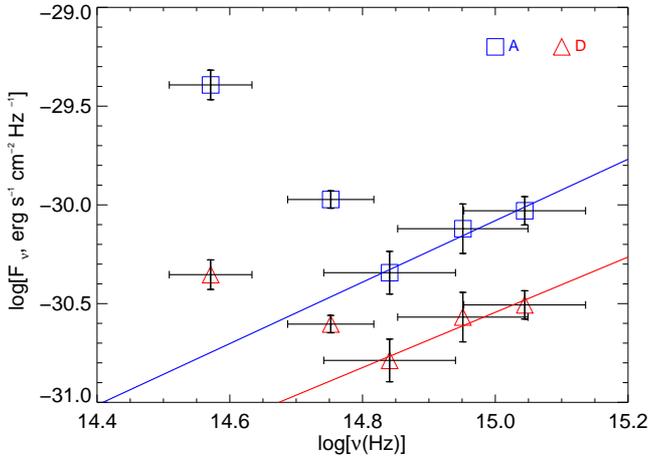}
		\caption{Spectral energy distributions of A (blue squares) and D (red triangles). The solid lines indicate a power-law model with a photon index $\alpha$ = 1.56$\pm$0.17 for A and 1.40$\pm$0.17 for D. There are excess in F555W and F814W bands. This could be correspond to what is expected from illumination effects of the nearby bright red star. Values of D are separated with -0.5 dex to clarify.}
		\label{Aksaker_fig4}
	\end{center}
\end{figure}

Spectral energy distributions (SEDs) for two candidates constructed from {\it HST} photometry are plotted in Figure \ref{Aksaker_fig4}. The SEDs (reddening corrected) for counterparts A and D are adequately fitted with simple power-law models with F$_{\nu}$ $\propto$ $\nu$$^{\alpha}$, where F$_{\nu}$ in units of erg cm$^{-2}$ s$^{-1}$ Hz$^{-1}$. The power-law indices ($\alpha$) are 1.56$\pm$0.17 for A and 1.40$\pm$0.17 for D. These fits are acceptable with $R^{2}$ of 0.96 and 0.94 where, $R^{2}$ is goodness-of-fit for a linear model. As aforementioned, we obtained a good fit by using only the F275W, F336W and F438W values to constrain the fits. We note that these indices are consistent with the most of the ULXs in the sample given by \cite{2011ApJ...737...81T} ($\alpha$ = -1 to 2). Such indices show the Rayleigh-Jeans tail of reprocessed disk emission with temperatures below about 6000 K \citep{2009MNRAS.392.1106G}.

\section{Results and Discussion}
\label{sec:res}
We have studied the optical properties of ULX X-1 in the galaxy NGC 2500 for the first time using the archival data from {\it HST}/WFC3/UVIS. Four possible optical counterparts were identified within $\sim0\farcs3$ error radius after the astrometric correction. However, two of them (B and C) were not investigated as possible optical counterparts of ULX X-1 due to their extended features and their absence in filters (F275W, F336W and F438W). Colors and absolute magnitudes ($M_{\rm V}$) of all possible candidates were obtained from the available data. It is noted that dereddened magnitudes of counterparts are quite faint when compared with the known ULXs optical counterparts. $M_{\rm V}$ values of these sources are consistent with the optical counterparts of ULXs in Table 4 of \cite{2011ApJ...737...81T}, which lie in the range -7 < $M_{\rm V}$ < -3. 
Assuming the optical light is dominated by the donor star, the spectral types of A and D were determined as early B type main sequence stars using the intrinsic colors and the absolute magnitudes in the Schmidt-Kaler Table \citep{1982lbg6.conf.....A}.

As given in Table \ref{tab2}, two possible optical counterparts (A and D)
have been detected in all images, making them the most noteworthy candidates. In their nearby environment any star group or cluster have not been observed within a radius of about 5$^{\prime\prime}$ region. We obtained a CMD ((U$-$B) versus B) to estimate the ages and masses of the candidates by using the mean extinction value as $E(U-B)=0.034$. The magnitudes and colors derived from {\it HST}/WFC3/UVIS data, compatible with those used in the available PARSEC isochrones and thus the derived  age and  mass values of A and D are almost the same as 45 Myr and 7 $M_{\rm \odot}$, respectively. The dereddened (U$-$B) color value is around -1 for both counterparts. It seems that counterparts A and D are not clearly bluer than the field stars.

\begin{figure}
	\begin{center}
		\includegraphics[width=\columnwidth]{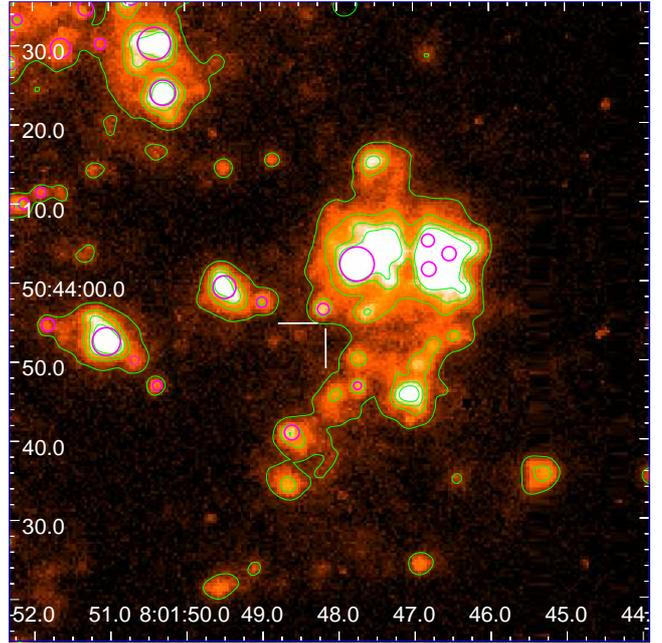}
		\caption{The NED $\mathrm{H}\alpha$ image of ULX X-1 and its environment. White lines (5$^{\prime\prime}$) show position of ULX X-1. Green contour levels are (88, 232, 376 and 520 ADU (1 ADU = 1.3$\times10^{-18}$ erg cm$^{-2}$ s$^{-1}$)) plotted on the image. The purple circles show HII regions from the catalog (see Section \ref{sec:res}). The north is up.}
		\label{Aksaker_fig5}
	\end{center}
\end{figure}

In order to examine possible nebulae and the star formation regions around ULX X-1 we have used the ground-based continuum subtracted $\mathrm{H}\alpha$ image (NGC$\_$2500$\_$I$\_$Ha$\_$d2009.fits) available at NED (NASA/IPAC Extragalactic Database). This image was obtained from 1.8 m The Vatican Advanced Technology Telescope (VATT) with an exposure time of 1800 s of NGC 2500. This image is given in Figure \ref{Aksaker_fig5}. Two bright regions are noticeable on the figure with a distance of 13$^{\prime\prime}$ $(1^{\prime\prime} = 48.4$ pc) to the position of X-1. Using $\mathrm{H}\alpha$ flux of the galaxy given as ${\log}  f =-11.60$ \citep{2008ApJS..178..247K} and the $\it{aper}$ function in IDL (Interactive Data Language) program, a total of 1.9$\times10^{6}$ counts (ADUs) were obtained from the $\mathrm{H}\alpha$ image for the entire galaxy. The errors of the counts are about 2$\%$ and 1 count corresponds to 1.3$\times10^{-18}$ erg cm$^{-2}$ s$^{-1}$. The contours are also plotted on $\mathrm{H}\alpha$ image. The source ULX X-1 falls on the contour value of 1.14$\times10^{-16}$ erg cm$^{-2}$ s$^{-1}$ which corresponds to a luminosity value of 1.40$\times10^{36}$ erg s$^{-1}$ at a distance of 10 Mpc. HII regions from the catalog of \cite{2015MNRAS.451.1307Z} are shown also on the image (Figure \ref{Aksaker_fig5}). An HII region is seen at about 1$^{\prime\prime}$ away from ULX X-1. The luminosity of this nearby HII region is given as 1.15$\times10^{37}$ erg s$^{-1}$ from catalog \citep{2015yCat..74511307Z}. This $\mathrm{H}\alpha$ luminosity is approximately ten times greater than the contour level the nearest to the ULX X-1. This decrease in $\mathrm{H}\alpha$ flux is clearly seen on the figure. Considering the nearby significant emission region, we think that a strong $\mathrm{H}\alpha$ emission cannot be originated from this source.

In many cases, there are possibilities of confusion between emissions from ULXs and background AGNs. Therefore, we also tried to discriminate such a confusion by estimating a ratio of X-ray to optical flux for A and D counterparts. Since simultaneous optical and X-ray observations do not exist, we used the available {\it Chandra} data (ObsID 7112) for X-ray and F555W data for optical. Assuming that the X-ray flux did not change at the time of F555W observation, the ratio was calculated by following formula log($F_{\mathrm{X}}/F_{\mathrm{opt}}$) $=$ $F_{\mathrm{X}}$ + m$_{V}$/2.5 + 5.37, where m$_{V}$ is the extinction corrected magnitude in F555W band and $F_{\mathrm{X}}$ is unabsorbed X-ray flux in 0.3$-$3.5 keV enery band. To determine $F_{\mathrm{X}}$ as 2.0$\times10^{-13}$ erg cm$^{-2}$ s$^{-1}$ , the Chandra PIMMS toolkit was used with an absorbed power-law index of 1.7 and avarage Galactic column density N$_{H}$$=$4.7$\times10^{20}$ cm$^{-2}$ (\citealp{1990ARA&A..28..215D}). The log($F_{\mathrm{X}}/F_{\mathrm{opt}}$) values for A and D were found to be 3.3 and 3.2, respectively. These are significantly greater than the accepted range between -1 to 1.7 for AGNs besides BL Lac objects, normal galaxies and normal stars (\citealp{1988ApJ...326..680M}; \citealp{1991ApJS...76..813S}). As we mentioned above, even though the bright red star's illumination effect on F555W (V) is present. we used m$_{V}$ value to calculate flux ratio. However, when considering the flux corresponding to the V wavelength from the power-law model the specified ratio is $\sim$ 3.6 and it still exceeds the upper limit for the give range. The log($F_{\mathrm{X}} / F_{\mathrm{opt}}$) value of our two sources are similar to other known ULXs's (\citealp{2008ApJ...675.1067F}; \citealp{2011ApJ...737...81T}; \citealp{2011ApJ...733..118Y}; \citealp{ 2016MNRAS.455L..91A,2016ApJ...828..105A}).

The optical emission from a ULX source might originate from its donor star or from the accretion disk around the compact object or both. The criterion to differentiate between X-ray binaries with high mass (HMXB) and low mass (LMXB) is defined by the formula $\xi= B_{\rm 0} + \log(F_{\rm \mathrm{X}})$
where  B$_{0}$ is the magnitude in B band and $F_{\mathrm{X}}$ is the 2$-$10 keV X-ray flux in units of $\mu$Jy the optical to X-ray fluxes. The $\xi$ values for HMXBs and LMXBs were found to be in the range of 12$-$18 and 21$-$22, respectively \citep{2017ARA&A..55..303K}. The calculated $\xi$ of the possible optical counterparts are about 20 i.e., closer to values of LMXBs. Other known ULXs also have $\xi$ values quite similar to LMXBs. Therefore, our analysis indicate that ULX X-1 has an emission mechanism similar to an LMXB for which the optical emission mostly comes from its accretion disk.

The resultant SEDs were fitted with power-law models. Best-fitting parameters of the model indicate that their optical emission is probably generated from reprocessing of the X-rays in the outer part of the disk \citep{2011ApJ...737...81T}. In our case, we do not have data for an optical variability check, which would provide a further support for the disk origin of the optical emission.

As a conclusion, we see that there are still several uncertainties in our understanding of the nature of ULXs. Therefore, future high-resolution observations in multiwavelenghts need to be encouraged for better interpretations.

\section*{Acknowledgements}
	We acknowledge support from the Scientific and Technological Research Council of Turkey (TUBITAK) through project no. 117F115.
	
\begin{table}
	\centering
	\caption{Log of {\it HST}/WFC3/UVIS observations of ULX X-1 in NGC 2500}
	\begin{tabular}{cccr}
		\hline\hline
		Filter & ObsID & Date & Exp. \\
		& & & (s) \\
		\hline
		F275W & ICDM38030 & 2013-08-28 & 2382 \\
		F336W & ICDM38040 & 2013-08-29 & 1119 \\
		F438W & ICDM38050 & 2013-08-29 &  965 \\
		F555W & ICDM38060 & 2013-08-29 & 1143 \\
		F814W & ICDM38070 & 2013-08-29 &  989 \\
		\hline\hline
		\label{tab1}
	\end{tabular}
\end{table}

\begin{table*}
	\caption{The properties of the possible optical counterparts of ULX X-1.} 
	\centering
	\small\addtolength{\tabcolsep}{-2pt}
	\begin{tabular}{cccccccccc}
		\hline \hline
		Source & R.A. & Dec. & F275W & F336W & F438W & F555W & F814W & Mv & $(B-V)_{0}$ \\
		\hline
		A & 8:01:48.11 & +50:43:54.58 & 24.95$\pm$0.18 & 25.64$\pm$0.31 & 26.75$\pm$0.27 & 26.26$\pm$0.11 & 25.72$\pm$0.19 & -3.74 & 0.48 \\
		B & 8:01:48.14 & +50:43:54.52 & - & - & - & 26.91$\pm$0.19 & 25.55$\pm$0.17 & -3.09 & - \\
		C & 8:01:48.14 & +50:43:54.61 & - & - & - & 27.07$\pm$0.21 & 25.22$\pm$0.13 & -2.93 & - \\
		D & 8:01:48.15 & +50:43:54.66 & 24.88$\pm$0.18 & 25.51$\pm$0.26 & 26.61$\pm$0.24 & 26.59$\pm$0.14 & 26.87$\pm$0.51 & -3.39 & -0.36 \\		
		\hline\hline
	\end{tabular}
	\label{tab2}
\end{table*}




\bibliographystyle{mnras}
\bibliography{bib} 


\bsp	
\label{lastpage}
\end{document}